\documentclass[aps,prresearch,twocolumn,amsmath,amssymb,nofootinbib,superscriptaddress,floatfix]{revtex4-2}

\usepackage{amsmath}
\usepackage{amssymb}
\usepackage{amsthm}
\usepackage{braket}

\usepackage[pdftex]{color} 
\usepackage{graphicx}
\usepackage{dcolumn} 
\usepackage{bm} 
\usepackage{float} 
\usepackage[driverfallback=dvipdfm]{hyperref} 
\usepackage{longtable}
\usepackage{ulem}   
\begin{document}

\author{Jian Wang}
\affiliation
{
School of Physics,
University of Science and Technology of China, Hefei, Anhui, 230026, China
}
\affiliation
{
International Center for Quantum Design of Functional Materials (ICQD), Hefei National Research Center for Interdisciplinary Sciences at the Microscale, University of Science and Technology of China, Hefei, Anhui 230026, China
}

\author{James Jun He}
\affiliation
{
International Center for Quantum Design of Functional Materials (ICQD), Hefei National Research Center for Interdisciplinary Sciences at the Microscale, University of Science and Technology of China, Hefei, Anhui 230026, China
}
\affiliation
{
Hefei National Laboratory, Hefei, Anhui 230088, China
}

\author{Qian Niu}
\affiliation
{
School of Physics,
University of Science and Technology of China, Hefei, Anhui, 230026, China
}
\affiliation
{
International Center for Quantum Design of Functional Materials (ICQD), Hefei National Research Center for Interdisciplinary Sciences at the Microscale, University of Science and Technology of China, Hefei, Anhui 230026, China
}

\title{ 
Space-time Supercrystals from Non-Abelian Electric Translation Symmetries
}

\begin{abstract}
Electronic supercrystals can form in spatial or temporal dimensions where traditional mechanisms usually require many-body interactions, such as Wigner crystals or discrete time crystals. We propose a novel approach for electronic supercrystals in 1+1D without requiring many-body interactions, but arising from the competition between characteristic space-time areas in periodically driven Su-Schrieffer-Heeger lattices under electric fields. Utilizing the non-Abelian dynamical symmetries described by the electric translation group, the area competition is characterized by fractal spectra and replicated bands, furnished with eigenstates crystallized in enlarged space-time unit cells forming space-time supercrystals under perturbations. We also report robust dynamical localization of electrons arising from the static topology of the SSH lattice. 
\end{abstract}

\date{\today}

\maketitle

\section{Introduction} 
The competition between characteristic physical scales in quantum materials often leads to novel quantum phenomena. A paradigmatic example is Hofstadter's butterfly—the fractal energy spectra observed in 2D lattices under magnetic fields, where the competition between the unit cell and the characteristic area occupied by a magnetic flux quantum induces replicated band structures characterized by the magnetic translation group symmetries\cite{zak1964,brown1964bloch}. Recently, Hofstadter quantum materials have emerge as an arena for exploring novel orders\cite{moller2015fractional,murthy2012hamiltonian,wang2020classification,Arbeitman2020,andrews2021stability,shaffer2021theory,shaffer2022unconventional}.

Space-time lattices are periodic driven quantum materials exhibiting both spatial and temporal translational symmetries, which have garnered significant attention due to their potential for showcasing quantum dynamical phenomena at nonequilibrium\cite{harper2020topology,rudner2020floquet}. Analogous to Hofstadter's butterfly, fractal spectra emerge in 1+1D space-time lattices under electric fields when the space-time unit cell competes with the characteristic space-time area occupied by an electric flux quantum. Initially noticed by Zhao et. al.\cite{zhao1995dynamic} in a one band model, these fractal spectra are named fractional Stark ladders. However, with the fundamental symmetry description still lacking, the dynamical aspects of fractional Stark ladders undergo limited exploration in existing literature\cite{holthaus1996localization,gluck2002wannier,niu1996atomic,zhao1996rabi}.

In this work, we utilize the electric translation group symmetries\cite{AshbyEM,ke2024} to explore area competition in space-time Su-Schrieffer-Heeger (SSH) lattices\cite{dal2015floquet,cheng2019observation,wu2020floquet} under electric fields and reveal a new dynamical mechanism for electronic supercrystals in spatial and temporal dimensions. Unlike traditional approaches relying on many-body interactions, our mechanism arises from the commensurability between the space-time lattice periodicity and electric flux quantum. Firstly, we derive a gauge invariant representation of the electric translations to solve for fractal spectra and replicated bands. We then classify the periodicity of the eigenstates using enlarged space-time unit cells characterized by the non-Abelian dynamical symmetries. Under perturbations preserving quasienergies or momenta, such eigenstates can be prepared and form space-time supercrystals exhibiting controllable spatial or temporal periods which are integer multiples of the lattice periods. Notably, this mechanism requires no many-body interactions and is distinguished from traditional electronic supercrystals in spatial or temporal dimensions, such as Wigner crystals\cite{Wigner34} and discrete time crystals\cite{else2016floquet,else2020discrete}. We also demonstrate that for commensurate area ratios, the symmetries protecting the topology of static SSH lattice evolve into a dynamical constraint enforcing electronic localization at arbitrary drive strength. Furthermore, the aforementioned phenomena are insensitive to the specific forms of drives, thus providing a comprehensive picture of quantum dynamics induced by space-time area competition.
\begin{figure}[htbp]
\centering
\includegraphics[width = .44\textwidth]{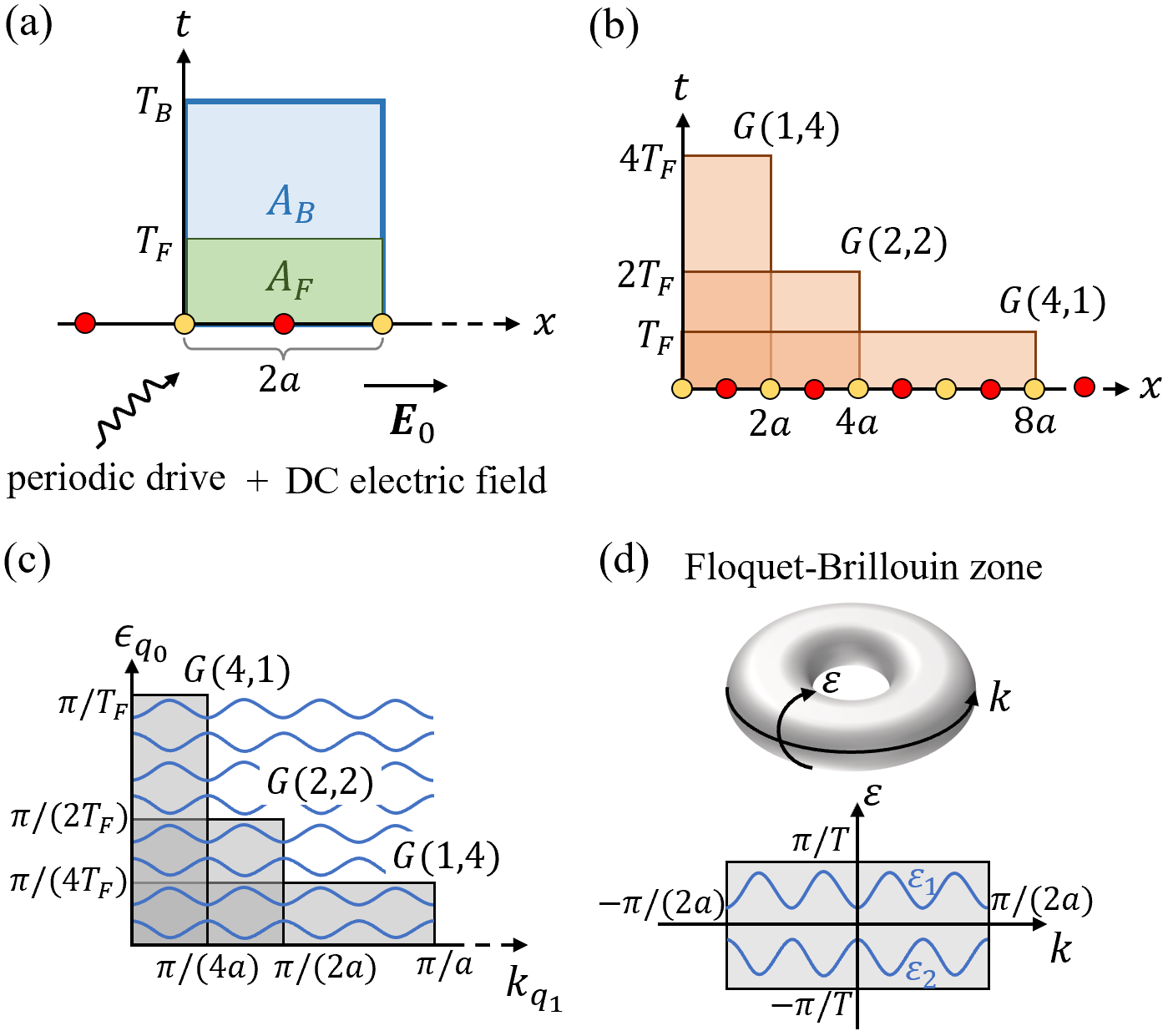}
\caption{(a) Illustration of the space-time area competition in the SSH lattice under periodic drive and DC electric field. The space-time unit cell $A_F = 2a\times T_F$ competes with the characteristic space-time area $A_B = 2a\times T_B$ occupied by an electric flux quantum, where $T_F$ is the period of the drive and $T_B$ is the period of Bloch oscillation. (b,c) Abelian subgroups $G(1,4),G(2,2)$ and $G(4,1)$ correspond to three different choices for the enlarged space-time unit cells presented in (b) and shrunk Floquet-Brillouin zones presented in (c) at $A_F/A_B=1/4$. (d) Illustration of the toric Floquet-Brillouin zone for $G(4,1)$. Electric quasienergy $\varepsilon$ is defined modulo $2\pi/T$ where $T=4T_F$, and momentum $k$ is defined module $\pi/a$. The dispersion relation $\varepsilon = \varepsilon_{\alpha}(k)$ gives rise to replicated quasienergy band structures.}
\label{fig: model and FBZ}
\end{figure}

\section{Model and Electric Translations}
To explore quantum dynamics under area competition, we focus on the spinless SSH model—a paradigmatic 1D system with tunable topology. Under a DC electric field $\bm{E}_0$, we adopt a gauge with solely the vector potential $\bm{A}(t) = -\bm{E}_0t$, which simplifies the analysis of dynamical symmetries by encoding the electric field effect into time-dependent Peierls phases. Using Peierls substitution, the effective tight-binding Hamiltonian is presented as follows:
\begin{eqnarray}
\label{eq: TB Hamiltonian 1}
H_0(t) = -\sum_{i}e^{i\omega_B t}(u a^\dagger_ib_{i+1}+v b^\dagger_ia_{i})+h.c.
\end{eqnarray}
where $\omega_B = eaE_0$,  $e$ denotes the elemental charge, $a$ indicates the distance between nearest neighbor sites, $u, v$
are the hopping elements and $a_i^\dagger, b_i^\dagger$ represent spinless electron creation operators on the two sublattices. The phase factor $e^{i\omega_B t}$ in the electron hopping terms preserves the translation symmetry at the expense of temporal independence. $T_B=2\pi/\omega_B$ is the period of Bloch oscillation\cite{bloch1929quantenmechanik,zener1934theory} and defines a characteristic space-time area $A_B = 2a\times T_B$, which is precisely the area occupied by an electric flux quantum by noting that $A_B=\phi_0/E_0 $ where $\phi_0 = 2\pi /e$ (we set $\hbar=1$) .

A competing characteristic space-time area is then introduced by applying Floquet drives. Our model thus describes 1+1D space-time SSH lattices under DC electric field (Figure \ref{fig: model and FBZ}a). We consider two types of drives: Case (I) is referred to as the optical drive, where $u$ and $v$ remain constant while an extra AC field is applied such that $\bm{E}(t) = [\bm{E}_0+\bm{E}_1\cos(\omega_F t)]$ denotes the total field. The corresponding Hamiltonian is given as
\begin{eqnarray}
\label{eq: TB Hamiltonian O}
 H_O(t) = -\sum_{i}e^{i[\omega_B t+\lambda\sin(\omega_F t)]}(ua^\dagger_ib_{i+1}+vb^\dagger_ia_{i})+h.c.
\end{eqnarray}
where $\lambda = ea E_1/\omega_F$. Case (II) is named the acoustic drive where the hopping elements oscillate as $u(t)=\bar u[1+\eta\cos(\omega_F t)], v(t)=\bar v[1+\eta\cos(\omega_F t)]$. The Hamiltonian for this case is written as 
\begin{eqnarray}
\label{eq: TB Hamiltonian A}
H_A(t)=-\sum_{i}e^{i\omega_B t}[u(t) a^\dagger_ib_{i+1}+v(t) b^\dagger_ia_{i}]+h.c.
\end{eqnarray}
Notably, light-induced Floquet engineering offers promising test beds for Case (I) in experiments\cite{bao2022light,zhou2023pseudospin,Zhou2023floquet}. The temporal period for both drives is given by $T_F = 2\pi/{\omega_F}$ which establishes a space-time unit cell of size $A_F = 2a\times T_F$. Importantly, our choice of gauge has introduced time dependent Peierls phases to describe the DC field such that Hamiltonian \ref{eq: TB Hamiltonian O} and \ref{eq: TB Hamiltonian A} do not exhibit the Floquet period $T_F$ when $A_B\neq A_F$. While the conventional temporal translation symmetry is lost in our gauge choice, the system retains gauge-invariant dynamical symmetries that combine translations with electric field effects.

These dynamical symmetries are encoded in the electric translation group, which is first introduced by Ashby and  Miller \cite{AshbyEM} in a periodic gauge for static lattices, we also note a recent article\cite{ke2024}. For our model, we present a gauge invariant representation of this group, whose generators are presented as follows:
\begin{subequations}
\label{eq: invariant generator}
\begin{align}
    & \hat T_0 = e^{T_F(\partial_t+ie \mathcal{A}_0)}
    \\
    & \hat T_1 = e^{2a(\partial_x+ie \mathcal{A}_1)}
\end{align}
\end{subequations}
where $\mathcal A_0$ and $\mathcal A_1$ are the components of the electromagnetic potential in a covariant dual gauge (see Supplemental Material \cite{supp} for detailed discussion on gauge invariance). In our gauge, specifically:
\begin{subequations}
\label{eq: generators}
\begin{align}
& \hat T_0  = e^{T_F\partial_t-i\pi r \hat x /a } \\
& \hat T_1 = e^{2a\partial_x}
\end{align}
\end{subequations}
where $\hat x$ is the position operator and $r = A_F/A_B$ is the area ratio. $\hat T_0$ represents a temporal translation by the drive period $T_F$, accompanied by a position-dependent phase modulation $e^{i\pi r \hat x/a}$ induced by accumulated electric flux. $\hat T_1$ corresponds to the conventional spatial translation by $2a$. Similar to the magnetic translation group\cite{brown1964bloch,zak1964} describing 2D lattices under magnetic fields, the electric translation generators are non-Abelian:
\begin{equation}
\label{eq: non-Abelian}
\hat T_0 \hat T_1 = e^{i2\pi r} \hat T_1 \hat T_0,
\end{equation}
reflecting the interplay between quantum dynamics and electromagnetic gauge invariance. Denoting the eigenvalue of $\hat T_0$ as $e^{i\epsilon T_F}$ and that of $\hat T_1$ as $e^{i k 2a}$, $\epsilon$ thus characterizes the average energy over a temporal translation by $T_F$ and $k$ is the Bloch momentum, both eigenvalues describe gauge invariant physical observables. Unfortunately, due to the non-Abelian commutation relation Equation \ref{eq: non-Abelian}, the vector $(\epsilon, k)$ is not a good quantum number when $A_B\neq A_F$ but is subjected to the uncertainty relation that $\epsilon$ and $k$ cannot be determined simultaneously. This uncertainty relation implies strong constraints on the dynamical behavior of our model. 

Notably, when $A_F$ and $A_B$ are commensurate and result in a rational area ratio $r=p/q$ where $p,q$ are co-primes integers, the electric translation group admits multiple Abelian subgroups providing good quantum numbers for our model. We denote a subgroup generated by $\hat T_0^{q_0}$ and $\hat T_1^{q_1}$ as $G(q_0,q_1)$, where $q_0$ and $q_1$ are integers satisfying $q_0 q_1 = q$, these generators are thus Abelian:
\begin{equation}
\hat T^{q_0}_0 \hat T^{q_1}_1 = (e^{i2\pi \frac{p}{q}})^{q_0q_1}  \hat T_1^{q_1}\hat T^{q_0}_0= \hat T_1^{q_1}\hat T^{q_0}_0,
\end{equation}
giving rise to an enlarged space-time unit cell with $q_0$ Floquet periods and $q_1$ spatial unit cells (Figure \ref{fig: model and FBZ}b). The space-time area of the enlarged unit cell equals to $2qaT_F$, which is the minimal space-time area where pseudo momenta and quasienergies can be determined simultaneously.

We thus denote the eigenvalue of $\hat T_0^{q_0}$ as $e^{i\epsilon_{q_0} q_0 T_F}$ and that of $\hat T_1^{q_1}$ as $e^{i k_{q_1} 2a}$ and define the electric quasi energy-momentum  $\bm{\kappa}_{q_0,q_1} = (\epsilon_{q_0},k_{q_1})$ which is a good quantum number presented in the shrunk Floquet-Brillouin zones of size $\pi/(q_1a)$ (Figure \ref{fig: model and FBZ}c). The Schrodinger equation thus determines the dispersion relation between $\epsilon_{q_0}$ and $k_{q_1}$.

\section{Dynamical Fractal Spectra}

The non-Abelian structure of the electric translation group imposes strong constraints on the dynamical spectra. When the space-time areas $A_F$ and $A_B$ are commensurate ($r=p/q$), the symmetry group admits Abelian subgroups $G(q_0,q_1)$ leading to fractal spectra and replicated quasienergy bands. To visualize this, we solve the eigen problem of Hamiltonian \ref{eq: TB Hamiltonian O} and \ref{eq: TB Hamiltonian A} under periodic boundary conditions. As shown in the previous section, there are multiple choices of good quantum number $\bm{\kappa}_{q_0,q_1}$ to label the eigenstates. In our gauge, the simplest choice is $\bm{\kappa}_{q,1}$ for subgroup $G(q,1)$, which is recognized as a conventional translation group generated by $\hat T_1$ and $\hat T_0^q = e^{T \partial_t}$, where $T=qT_F = pT_B$. Denoting $\varepsilon = \epsilon_{q}$, according to Floquet-Bloch band theory\cite{gomezleon2013}, the dispersion relation reads $\varepsilon =\varepsilon_\alpha(k)$, where $\alpha=1$ or $2$ is the band index. The eigenstates are presented as 
$
\label{eq: dynamical eigenstates}
\psi_{\alpha,k}(x,t) =e^{i[kx-\varepsilon_\alpha(k) t]}u_{\alpha,k}(x,t),  
$
which form a complete basis at non-equilibrium with $u_{\alpha,k}(x,t)=u_{\alpha,k}(x+2a,t)=u_{\alpha,k}(x,t+T)$. The electric quasienergy $\varepsilon_\alpha(k)$ thus describes the average energy of $\psi_{\alpha,k}$ over $q$ Floquet periods, giving rise to band structures in a toric Floquet-Brillouin zone (Figure \ref{fig: model and FBZ}d).

\begin{figure}[htbp]
\centering
\includegraphics[width = .48\textwidth]{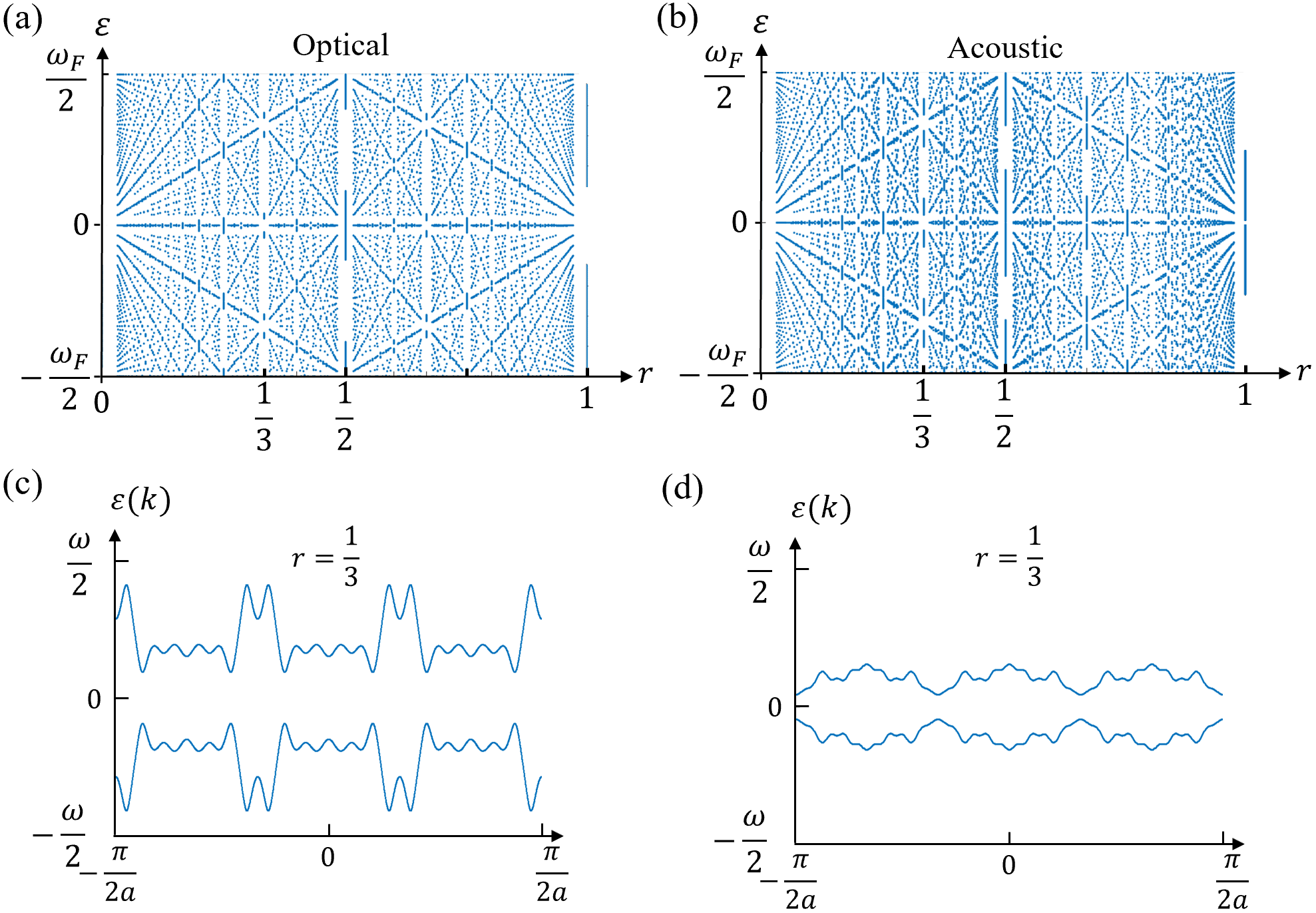}
\caption{(a) Fractal spectra at $r=p/q$ for the optical drive at hopping elements $u=3\omega_F$, $v=4.2\omega_F$ and drive strength $\lambda=2.5\omega_F$ where $\omega_F$ is the frequency of the drive and $\omega = \omega_F/q$.
(b) Fractal spectra for the acoustic drive at $u=2\omega_F$, $v=3.2\omega_F$ and $\eta=1.2$. (c) Band structure of (a) at $r=1/3$. (d) Band structure of (b) at $r=1/3$. The electric quasienergy $\varepsilon$ are $q$-fold degenerate due to replicated band structures.}
\label{fig: fractal spectra}
\end{figure}

To solve for the dispersion relation, we define the Lagrangian operator\footnote{This name originates from path integral, where the expectation value of $L = i\partial_t-H$ over coherent state defines the path integral Lagrangian.} as $
L = i\partial_t - H(t)$, the functions $u_{\alpha,k}$ are thus eigenfunctions of $L$ with eigenvalues $-\varepsilon_\alpha(k)$\cite{gomezleon2013}. Therefore, the dynamical spectra of Hamiltonian \ref{eq: TB Hamiltonian O} and \ref{eq: TB Hamiltonian A} can be obtained by seeking the eigenvalues of $L$ in the space spanned by $u_{\alpha,k}$. This is accomplished by performing Fourier transformations on the Lagrangian operators with respect to the spatiotemporal periodicity of $u_{\alpha,k}$ and diagonalizing the resulting matrices\cite{rudner2020floquet}: 
\begin{eqnarray}
    &&L_{O(A)} = \sum_{m,n=-\infty}^{\infty}
    \begin{pmatrix}
    a^\dagger_{k,m} & b^\dagger_{k,m}
    \end{pmatrix}
    L_{O(A)}^{mn}
    \begin{pmatrix}
    a_{k,n} \\ 
    b_{k,n}
    \end{pmatrix}.
\end{eqnarray}
The expressions for $L^{mn}_{O(A)}$ are presented in the Supplemental Material\cite{supp}. The eigensystem of the above matrix thereby yields $\varepsilon_\alpha(k)$ and $u_{\alpha,k}$.  

Notice that the area competition shrinks the length of the Floquet-Brillouin zone to $\omega=\omega_F/q$ in the $\varepsilon$ direction such that $\varepsilon\in(-\omega_F/(2q),\omega_F/(2q)]$ . Consequentially, there are $q$ copies of the two quasienergy bands $\varepsilon_1(k)$ and $\varepsilon_2(k)$ within a fixed energy range $(-\omega_F/2,\omega_F/2]$ containing $q$ replicas of the Floquet-Brillouin zone. The resulting dynamical spectra thus exhibit a fractal dependence on the area ratio $r$ as shown in Figure \ref{fig: fractal spectra}a and b\cite{zhao1995dynamic}. This fractality is also comprehended by perturbing the bands at $r$ with an additional DC field $\delta E_0$, which shifts the area ratio to $r+ \delta E_0ea/\omega_F$ and turns each band into a set of Wannier-Stark ladders\cite{wannier1962dynamics} with level spacing $\delta \varepsilon \sim\delta E_0ea$. These levels become dense when $\delta E_0$ is infinitesimal and give rise to fractal spectra.   

Interestingly, as depicted in Figure \ref{fig: fractal spectra}c and d, the quasienergies exhibit $q$-fold degeneracy, with the bands replicating $q$ times along the $k$ axis. This is a consequence of the symmetry $\hat T_0$. Denoting an eigenstate with quasienergy $\varepsilon_\alpha(k)$ as $\ket{\psi_{\alpha,k}}$, it is demonstrated that   
\begin{eqnarray}
&&\hat{T_1}\hat{T_0}\ket{\psi_{\alpha,k}}
\nonumber\\=&&e^{i2\pi\frac{p}{q}}\hat{T_0}\hat{T_1}\ket{\psi_{\alpha,k}}
\nonumber\\=&&e^{i(\frac{\pi p}{qa}+k)(2a)}\hat{T_0}\ket{\psi_{\alpha,k}}.
\end{eqnarray}
Consequentially, $\hat{T^n_0}\ket{\psi_{\alpha,k}}$ with $n=0,...,q-1$ are degenerate eigenstates with $\varepsilon_\alpha(k)=\varepsilon_\alpha[k+np\pi/(qa)]$, resulting in replicated bands. This feature bears resemblance to the degeneracy of Hofstadter bands\cite{azbel1964energy,Hofstadter1976} arising from the magnetic translation symmetries. Furthermore, the non-Abelian electric translations enforce unique dynamical behavior of the electronic states in our model: extended spatial or temporal periodicity described by enlarged space-time unit cells, giving rise to space-time supercrystals.

\section{space-time supercrystals}

Electronic states can extend the spatial or temporal periods of the static or driven lattices and form supercrystals. Traditional mechanisms require many-body interactions, such as Wigner crystals in lattice materials\cite{Wigner34,smolenski2021signatures,zhou2021bilayer,falson2022competing} and discrete time crystals\cite{else2016floquet,else2020discrete}. In our model, electronic supercrystals can arise from the space-time area competition between the lattice periodicity and electric flux quantum, without requiring many-body interactions. In the following, we reveal this mechanism by classifying the periodicity of the eigenstates using the electric translation subgroups $G(q_0,q_1)$.

The non-Abelian structure of the electric translation group gives rise to an intriguing symmetry constraint: although the Hamiltonians admit all the electric translation symmetries, the single-electron states exhibit only an Abelian subgroup of these symmetries and break the rest, otherwise the uncertainty relation between $\epsilon$ and $k$ will be violated. At commensurate area ratio $r=p/q$, this constraint forces the single-electron states to crystallize in enlarged space-time unit cells described by Abelian subgroups $G(q_0,q_1)$, which extend the spatial and temporal period to $q_0T_F$ and $2q_1a$ (Figure \ref{fig: model and FBZ}c), forming space-time supercrystals.  

\begin{figure}[htbp]
\centering
\includegraphics[width = .48\textwidth]{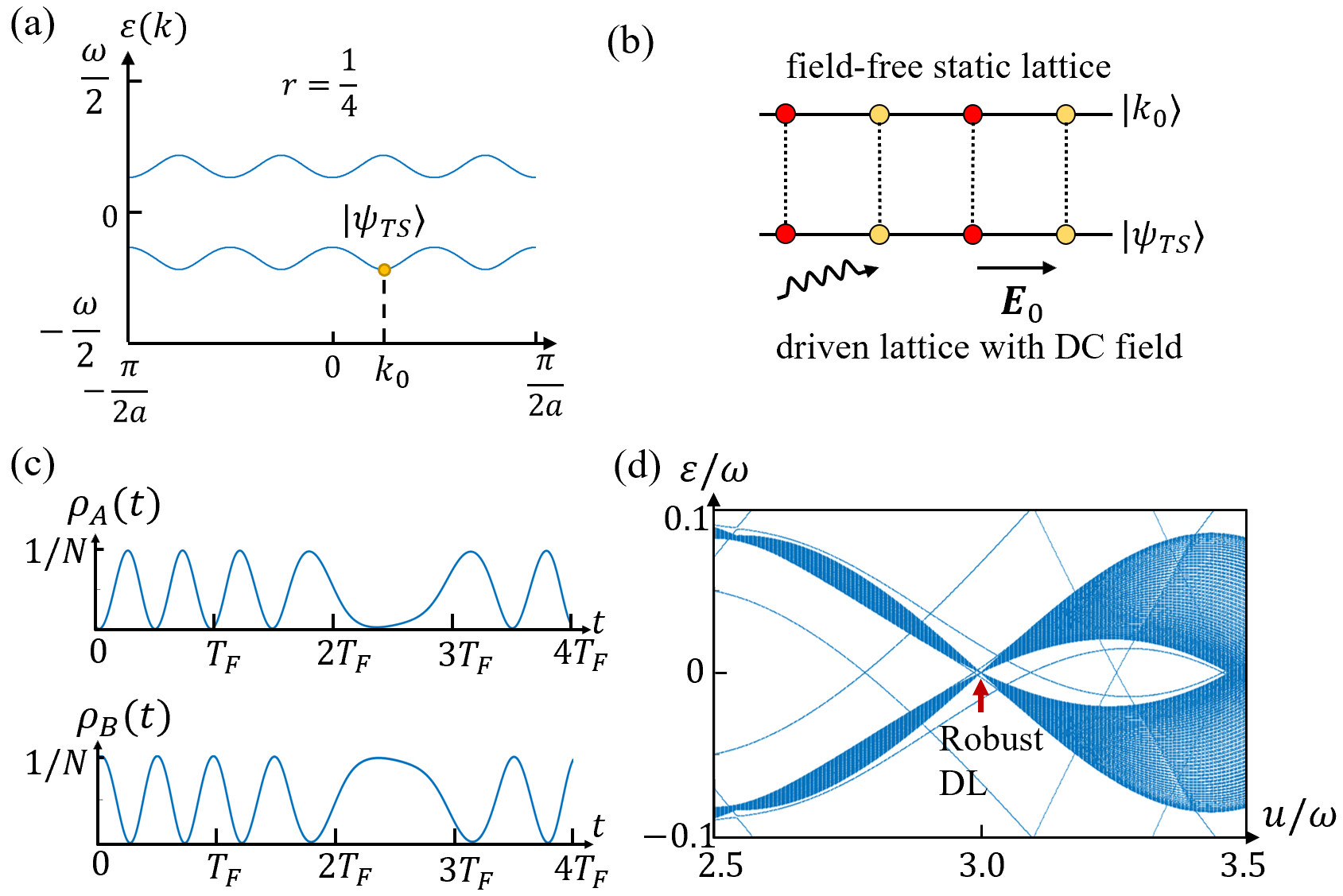}
\caption{
(a) Illustration of the replicated bands of subgroup $G(4,1)$ at $r=1/4, u=\omega_F, v=1.3\omega_F$ and $\lambda = 0.05\omega_F$ where a time supercrystal state $\ket{\psi_{\mathrm{TS}}}$ is excited by the perturbation preserving momentum $k=k_0=\pi/(8a)$. (b) Illustration of the approach to prepare $\ket{\psi_{TS}}$. The static SSH lattice in Bloch state $\ket{k_0}$ couples to the space-time SSH lattice weakly and excites time supercrystal state $\ket{\psi_{\mathrm{TS}}}$. (c) Evolution of the probability densities of $\ket{\psi_{\mathrm{TS}}}$ on sublattice A and B. $N$ is the number of unit cells. The evolution repeats after the period $4T_F$, which identifies $\ket{\psi_{\mathrm{TS}}}$ as a time supercrystal.  (d) Band collapse for the optical drive by modulating $u$ at $r=1/4, v=3\omega, \lambda=0.125\omega$. At any rational $r$, the band collapse point at $u=v$ exists for arbitrary $\lambda$, indicating robust dynamical localization (DL) of bulk electrons without requiring fine tuning. Edge modes are also presented. }
\label{fig: parameter detuned}
\end{figure}

The eigenstates of subgroup $G(q,1)$ discussed in the previous section are recognized as time supercrystals. Figure \ref{fig: parameter detuned}a demonstrates an example $\ket{\psi_{\mathrm{TS}}}$ from the replicated bands at $r=1/4$ under optical drive. This state preserves momentum $k=k_0$ and exhibits spatial period $2a$. Remarkably, the probability densities of this state $\rho_A(t)$ and $\rho_B(t)$ on sublattice $A$ and $B$ evolve with temporal period $4T_F$ which extends the temporal period of the Floquet drive (Figure \ref{fig: parameter detuned}c). Our numerical results reveal rich time supercrystals with temporal period $qT_F$ at $r=p/q$ in the bulk and the boundaries of our model\cite{supp}.

It is natural to ask how such time supercrystals can be prepared without relying on many-body interactions. To answer this question, we propose a simplified scheme for theoretical clarity, which may inspire experimental implementation. By noting that $k$ is a good quantum number that distinguish $\ket{\psi_{\mathrm{TS}}}$ from the degenerate eigenstates, we introduce a perturbation preserving $k=k_0$ by coupling the space-time lattice to a static SSH lattice at the Bloch state $\ket{k_0}$ (Figure \ref{fig: parameter detuned}b). If the coupling strength is weak enough, this perturbation will excite $\ket{\psi_{\mathrm{TS}}}$ at $k=k_0$ without altering the band structures. Note that our approach preserves the periodicity of the space-time SSH lattice and involves no many-body interactions.

Similarly, spatial supercrystals can arise from the eigenstates of the subgroup $G(1,q)$, where the spatial period is extended to $2qa$ and the temporal period remains $T_F$. Such eigenstates can be prepared using perturbations preserving the quasienergies in one Floquet period. Furthermore, when $q=q_1q_2$ is a composite number where $q_1>1$, the eigenstates of $G(q_1, q_2)$ extend both the spatial and temporal period of the space-time lattice and form space-time supercrystals. Either momentum-preserving or quasienergy-preserving perturbations can prepare such supercrystals. Examples of spatial and space-time supercrystals are presented in the Supplemental Material\cite{supp}. Importantly, the periodicity of the electronic supercrystals in our model is controllable by detuning the DC field strength or the Floquet period. 

\section{robust dynamical localization}
Space-time lattices possess a diverse parameter space encompassing lattice hopping elements and drive strengths. By modulating these parameters for our model, its dynamical behavior can change significantly.  At zero drive strength, our model exhibits Wannier-Stark localization\cite{wannier1962dynamics} due to the DC field, while when the drives are applied, the bulk electrons commence to delocalize\cite{ivanov2008coherent,ray2018drive,ringot2000experimental} except for certain values of the parameters. These localization-delocalization transitions can be described by the quasi-energy bands: when the band widths vanish, the circumstance is referred to as band collapse and indicates the dynamical localization of bulk electrons\cite{dunlap1986dynamic,holthaus1992collapse}.

Notably, at the topological critical point of the static SSH lattice\cite{SSH} where $u=v$, the bands of our space-time lattice model collapse to zero energy at any rational $r$ and all values of the driving strengths $\lambda$ and $\eta$, giving rise to robust dynamical localization of electrons without requiring fine tuning (Figure \ref{fig: parameter detuned}a). We prove in the following that this dynamical constraint originates from the symmetries protecting the static topology of the SSH lattice—sublattice symmetry $\mathcal S$ and the unitary inversion symmetry $\mathcal I$, enforcing invariant evolution under $\hat T^q_0$.

Consider the time-dependent Bloch Hamiltonian for the optical drive: 
\begin{eqnarray}
    H_{O}(k,t) = e^{i[\Theta(t)-ka]}\begin{pmatrix}
0& u \\
v & 0
\end{pmatrix} +h.c.
\end{eqnarray}
where $\Theta(t)=\omega_B t+\lambda\sin(\omega_F t)$, we denote the two quasi-energy bands as $\varepsilon_1(k)$ and $\varepsilon_2(k)$. The sublattice symmetry $\mathcal S$ is thus represented by $\sigma_z$ and the unitary inversion symmetry $\mathcal{I}$ by $\sigma_x$.  Notice that $\{\sigma_z, H_O(k,t)\}=0$ such that $\varepsilon_1(k) = -\varepsilon_2(k)$\cite{harper2020topology}. Furthermore, $\mathcal I$ conserves momentum at $u=v$ because $[\sigma_x, H_O(k,t)]=0$ and implies that $\varepsilon_1(k) = \varepsilon_2(k)$, therefore $\varepsilon_1(k) = \varepsilon_2(k) = 0$. According to Floquet-Bloch theorem\cite{gomezleon2013}, this equation enforces an invariant evolution operator $U(k,t)$ under $\hat{T}_0^q$ where $U(k,t) = \hat{T}_0^q U(k,t) \hat{T}_0^{-q} = U(k,t+T)$. Consequentially, the average motion of bulk electrons over $T$ vanishes. This is precisely the physical picture of dynamical localization. The proof for the acoustic drive is similar. The static topological critical point $u=v$ thus characterizes a dynamical localization point at arbitrary drive strengths.

\section{Conclusions}
In summary, we have derived the gauge invariant representation of the electric translation group to describe periodically driven lattices under electric fields. Upon the competition between space-time unit cell $A_F$ and the characteristic space-time area $A_B$ occupied by a flux quantum, this group encodes non-Abelian dynamical symmetries. Employing this group, we have investigated space-time SSH lattices in DC electric fields. When the area ratio $r=A_F/A_B$ is rational, we reveal fractal spectra and replicated bands for our model. The non-Abelian dynamical symmetries further characterize enlarged space-time unit cells for the eigenstates, giving rise to space-time supercrystals of electrons under perturbations without requiring many-body interactions. We also show that at rational $r$ the symmetries protecting the static topology of the SSH lattice give rise to a dynamical constraint forcing the spectra to collapse at balanced hopping elements, such that the electrons localize at arbitrary drive strength. These phenomena are explored for both optical and acoustic drives, thus providing a comprehensive picture of the quantum dynamics of space-time lattices under electric fields.

Our work unveils some intriguing avenues: 1.The optical drive (Case I) may be implemented in solid materials where Floquet engineering enables precise control of drive frequency modulations\cite{bao2022light,zhou2023pseudospin,Zhou2023floquet}. Probing the fractal spectra via time-resolved spectroscopy or transport measurements would validate our prediction for space-time supercrystals; 2. The robust dynamical localization at $u=v$ suggests a pathway to design noise resistant quantum devices, leveraging the interplay between dynamical symmetries and static topology; 3. Generalizing the electric translation group to higher dimensional systems with complex electromagnetic fields may uncover novel phases, such as space-time analogs of quantum Hall phases.
\section{acknowledgments}
We thank Haoshu Li for useful discussions. The authors are supported by the National Natural Science Foundation of China (Grant No. 12234017) and
the National Key Research and Development Program of China (Grant No. 2023YFA1406300). J.W. is also supported by the China Postdoctoral Science Foundation (Grant No. 2023M733415).


\bibliography{bibliography}

\end{document}